\documentclass[sigconf]{acmart}
\setcopyright{iw3c2w3}

\AtBeginDocument{%
  \providecommand\BibTeX{{%
    \normalfont B\kern-0.5em{\scshape i\kern-0.25em b}\kern-0.8em\TeX}}}

\copyrightyear{2021}
\acmYear{2021}

\acmConference[ORSUM '21]{ORSUM '21: 4th Workshop on Online Recommender Systems and User Modeling at ACM RecSys 2021}{October 02, 2021}{Online}
\acmBooktitle{ORSUM '21: 4th Workshop on Online Recommender Systems and User Modeling at ACM RecSys 2021, October 02, 2021, Online}

\begin{document}

\title{PEEK: A Large Dataset of Learner Engagement\\with Educational Videos}


\author{Sahan Bulathwela$^1$, Mar\'ia P\'erez-Ortiz$^1$, Erik Novak$^2$, Emine Yilmaz$^1$ and John Shawe-Taylor$^1$}
\email{m.bulathwela@ucl.ac.uk}
\affiliation{%
 \institution{$^1\ $Centre for Artificial Intelligence, University College London (UK)}
}

\affiliation{%
 \institution{$^2\ $Jožef Stefan Institute, Jožef Stefan International Postgraduate School (Slovenia)}
}


 

\renewcommand{\shortauthors}{Bulathwela, et al.}

\begin{abstract}
%
 In this work, we release a large and novel dataset of learners engaging with educational videos in-the-wild. The dataset, named Personalised Educational Engagement with Knowledge Topics (\emph{PEEK}), is one of the first publicly available datasets that address personalised educational engagement.
Educational recommenders have received much less attention in comparison to e-commerce and entertainment-related recommenders, even though efficient personalised learning systems could improve learning gains significantly.
One of the main challenges in advancing this research direction is the scarcity of large, publicly available datasets. In the PEEK dataset, educational video lectures have been associated with Wikipedia concepts related to the material of the lecture, thus providing a humanly intuitive taxonomy. We believe that granular learner engagement signals, in unison with rich content representations, will pave the way to building powerful personalisation algorithms that will revolutionise educational and informational recommendation systems. Towards this goal, we 1) construct a novel dataset from a popular video lecture repository, 2) identify a set of benchmark algorithms to model engagement, and 3) run extensive experimentation on the PEEK  dataset to demonstrate its value. Our experiments with the dataset show promise in building powerful informational recommender systems. The dataset and the support code is available at \url{https://github.com/sahanbull/PEEK-Dataset}.

\end{abstract}



\maketitle

\section{Introduction}
Developing artificial intelligence systems that, mildly at least, understand the structure of knowledge is
foundational to
building effective recommendation systems for lifelong education\cite{pardos2014affective,trueeducation,yu2020mooccube}, as well as for many other applications related to knowledge management and tracing\cite{lewis2020retrieval,yano2016taking}. In the context of personalised education, the research communities have tirelessly worked on building Intelligent Tutoring Systems (ITS) which have been their focus from the early applications of AI in education\cite{corbett1994knowledge}. ITS focus heavily on personalising testing opportunities that will allow learners to demonstrate their mastery of a Knowledge Component (KC), an atomic unit of knowledge that can be learned and mastered. These systems are designed for learning experiences with limited scope (such as learning about a specific topic, short course, etc.) which makes hand crafting KCs and encouraging repetitive test taking operationally feasible. More recent developments in online education has led to the boom of Open Educational Resources (OERs) and Massively Open Online Courses (MOOCs) moves from ITS to educational recommendation systems and personalised e-learning platforms that can cater a more informal learning setting where self motivated learners discover educational materials online to pursue their lifelong learning goals. To succeed in this landscape, a wider range of factors such as the diversity of  learners, their personal drivers and how these drivers change over time has to be accounted for. One of the major barriers in building next generation educational recommendation systems is the scarcity of publicly available datasets. 

\subsection{Our Contribution} 

We publish \emph{(P)ersonalised (E)ducational (E)ngagement Linked to (K)nowledge Topics} (\texttt{PEEK}), a novel dataset that comprises of watch time interactions of over 20,000 informal learners watching over 10,200 unique educational video lectures in an OER repository.  Our contribution is two-fold, consisting of: i) constructing a novel dataset to predict learner engagement with educational videos and ii) formalising a prediction task with several baselines. The video lectures are partitioned into multiple fragments where the parts are transcribed and associated with Wikipedia topics using entity linking. PEEK  dataset uses a humanly intuitive content representation where the topics are atomic Wikipedia pages. Having humanly intuitive content representations allows building explainable learner models that encourage learner meta-cognition and self-regulation, a valuable feature of a technology enhanced learning system\cite{bull2008metacognition}. The normalised watch time of individual users is discretised and provided as a target label defining a binary (engaged vs. not engaged)  prediction task. We further identify several baselines from prior work and establish a formal task to benchmark predictive performance on the proposed dataset.             

\section{Related Work} \label{lit_review}

Publication of this dataset is inspired by advances in multiple research verticals. PEEK dataset is an educational recommendation dataset that contains learner interactions with fragments of video lectures. We outline the relevant works from these different research verticals in this section.

\subsection{Intelligent Tutoring Systems and Educational Recommenders}

\emph{Knowledge Tracing} (KT) and \emph{Item Response Theory} (IRT), the foundational methods used in Intelligent Tutoring Systems, are evolving from traditional machine learning\cite{corbett1994knowledge,Yudelson13,vie2019knowledge,PFA_model,Pelanek2017}  to deep learning models\cite{deep_kt,goal_based_edrec} that improve predictive performance by sacrificing interpretability and transparency, crucial requirements for many of these systems\cite{barria2019making,bull2020there}. It is also hard to overlook the data hunger of neural approaches considering the possible data scarcity in an educational setting where additional information should be exploited. While accompanying these additional drawbacks, deep learning models used in these tasks are still not guaranteed to outperform classical approaches\cite{wilson2016back}. 

Another major drawback of ITS research is its primary focus into predicting knowledge mastery via explicit learner feedback that comes through test taking. Although test taking is a reliable approach to verify learning, this approach comes with its fair share of limitations. Tests need to be carefully crafted by experts in a way that the skill mastery of learners on specific skills can be carefully verified with accompanying hints\cite{fyfe2016providing,assist_math1,choi2020ednet} costing substantial human capital. Scaling automatic question generation is an open challenge. The other stakeholder segment put into pressure by question solving is the learners themselves. Putting the learner in front of tests too frequently can significantly hinder the user experience driving them to abandon the system.

Beyond the use of costly explicit user interactions,
there are various other implicit interactions learners execute within an interactive e-learning system that can be potentially exploited. Learner engagement with educational materials is also indicated by implicit interactions such as watching/pausing/rewinding/replaying video lecture, taking part in discussion fora etc.\cite{ramesh2014learning}. In the context of videos, although the potential of using watch time for engagement prediction\cite{beyondviews} and personalisation\cite{Covington2016} has led to positive results, this idea is under-explored in the education domain with only a handful of studies showing promise\cite{Guo_vid_prod,context_agnostic_engagement} in population-based engagement prediction. Although there is a strong interest in building state-aware, personalised educational systems in the recent years\cite{sigir_sum_20,sum_20}, an obvious reason for slow growth of personalised educational recommenders is the scarcity of publicly available datasets. 

Unfortunately, there is very limited work in the Information Retrieval (IR) community that deals with a personalisation scenario similar to PEEK dataset as IR scenarios in education are significantly different from general click behaviours for which datasets are abundantly available. TrueLearn Novel model\cite{truelearn} is the only recent algorithmic contribution that deals with a similar dataset where topical contents automatically extracted from educational videos are used to predict learner engagement. 
Apart from sophisticated learner modelling techniques such as TrueLearn Novel, other basic recommendation strategies, such as content-based filtering and collaborative filtering, are also applicable to this dataset\cite{katz2011using}.

\subsection{Video Fragments}

Although majority of studies in the field focus on recommending content items that are relevant to a learner, the possibility of recommending parts of items (e.g. a part of a video in contrast to an entire video) has been investigated lately. Proximity-aware information retrieval that exploits the positional structure of tokens is not a new idea\cite{schenkel2007efficient} in IR. Segmenting videos and building a table of contents using video segments has proven to be useful in efficiently summarising video content\cite{videoken,google_chapters}. Breaking informational videos into fragments has also shown promise in efficient previewing\cite{x5learn,x5learn2,chen2018temporally} and enabling non-linear consumption of videos\cite{non_lin}. Our prior proposal, TrueLearn Novel\cite{truelearn} model demonstrates the potential of using fragment-wise recommendation in educational recommenders. For these experiments, we created video parts of 5,000 characters (5 minutes) in the fragmentation process. This allows the e-learning system to have video fragments that contain a satisfactory amount of knowledge while keeping the video fragment length at a favourable value in terms of retaining viewer engagement\cite{Guo_vid_prod}. While TrueLearn Novel demonstrates promise, we strongly believe that availability of such a dataset to the public is critical in pushing the frontiers of research in (fragment-based) educational recommendation. PEEK dataset addresses this need. 

\subsection{Scalable Feature Extraction}

In addition to manual question generation, systems using KT and IRT often relies on expert labelling of the \textbf{Knowledge Components (KCs)}\cite{assistments_data} (sometimes also for the hierarchy of knowledge\cite{bauman2018recommending} and defining a Q-matrix\cite{tatsuoka1983rule}), which is time consuming and not scalable to lifelong learning applications. There are various machine learning based unsupervised learning approaches that are used to extract latent topics from textual content. Approaches such as Latent Dirichlet Allocation (LDA)\cite{blei2003latent}, Latent Semantic Analysis (LSA)\cite{dumais2004latent} and other probabilistic approaches\cite{hofmann2013probabilistic,liang2019collaborative} are potential candidates in this area. However, these unsupervised learning approaches suffer from complex hyper-parameter tuning processes and limited interpretability of the discovered \emph{latent} KCs creating gaps in transparency. A step forward in this area has been using \emph{explicit} representations based on taxonomies such as Wikipedia concepts\cite{gabrilovich2007computing,egozi2011concept}.    

\emph{Wikification}, a form of entity linking\cite{wikifier,tagme} 
has shown great promise for automatically capturing the KCs covered in an educational resource. This technology is on a promising path towards providing \emph{automatic}, \emph{humanly-intuitive (symbolic)} representations from Wikipedia, representing at the same time \emph{up-to-date knowledge} about \emph{many domains}. Wikipedia, being one of the largest encyclopedias in the world, evolves with time due to the contributor population that constantly updates it. Due to these reasons, we use Wikification\cite{wikifier} to generate KCs that are included in the PEEK dataset.

\subsection{Related Datasets}

The majority of datasets that are related to this task come from the \emph{knowledge tracing} domain\cite{corbett1994knowledge,deep_kt} which focuses on recovering knowledge/skill mastery of learners based on how they answer to specific exercises. This approach, as discussed in Section \ref{lit_review}, is much more explicit and effort intense than inferring it using implicit feedback (e.g. watched patterns of educational videos). Majority of the research in this domain has been based on ASSISTments data\cite{assistments_data}, that records a series of learners solving tests in an ITS platform. These datasets are heavily biased towards mathematics knowledge as the ASSISTments tool was used for mathematics education from the beginning\cite{assist_math1,mcguire2017counterintuitive}.

A few publicly available datasets to solve knowledge tracing problem exist. They include mathematics ASSISTments\footnote{\url{http://www.assistmentstestbed.org/}}\cite{assistmentHurstC18,walkington2019effect}, problem solving interactions\cite{choi2020ednet}, or multiple choice questions\footnote{\url{https://www.microsoft.com/en-us/research/event/diagnostic-questions-neurips2020/}}\cite{wang2020diagnostic,wang2021educational}), all of which are noteworthy although they do not focus on implicit feedback and engagement. MOOCCube is a very recently released dataset that contains a spectrum of different statistics relating to learner-MOOC interactions including implicit and explicit test taking activity\cite{yu2020mooccube}. Although this dataset may contain data that can be used to predict learner engagement with implicit feedback, the dataset has only been used in prerequisite detection task which is very different. However, our prior experiments have consistently demonstrated that engagement prediction via online learning\cite{truelearn} can be achieved with a dataset that is similar to PEEK in structure.  

Although previous work based on popular video repositories such as edX\cite{Guo_vid_prod}, Khan Academy\cite{khan_bigdata}, VideoLectures\cite{truelearn,trueeducation} and YouTube\cite{beyondviews,Covington2016} has evidenced the existence of datasets that include implicit engagement signals, these datasets are never made public due to their proprietary nature. 
The scarcity of large scale publicly available datasets for predicting learner engagement with educational videos constrain the growth of the field. In this context, the Personalised Educational Engagement (\emph{PEEK}) dataset, is the first and largest learner video engagement dataset that will be publicly released with humanly-interpretable Wikipedia concepts and the concept coverage associated with the video lecture fragments.
  
\section{\texttt{PEEK} Dataset}

\begin{figure*}[ht]
\begin{center}
    \centerline{\includegraphics[width=.9\linewidth]{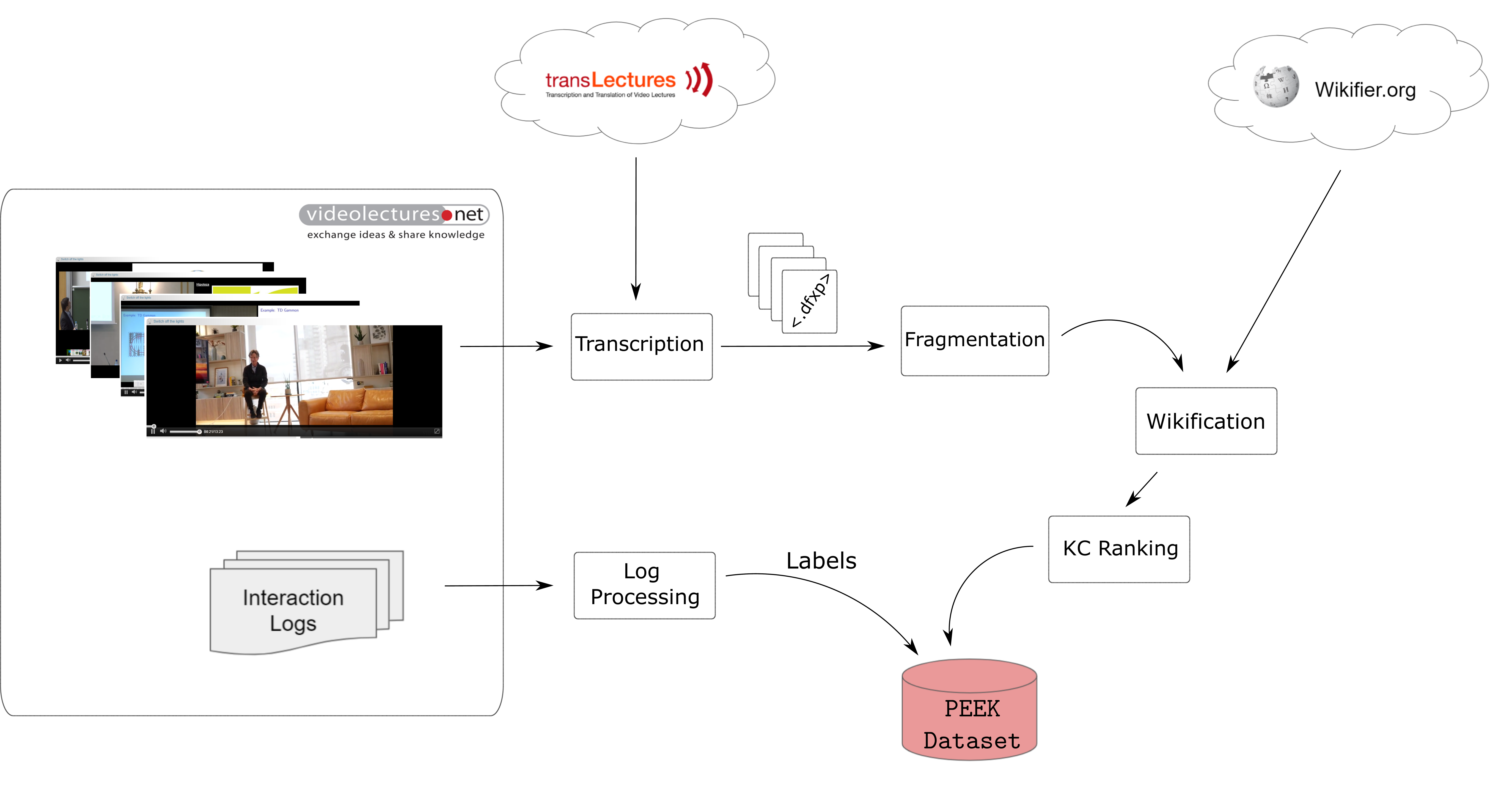}}
    \caption{The video data and the learner interaction logs from VLN repository are processed separately to create the Wikipedia-based KCs and also the discrete engagement signals that are published in PEEK dataset.}
    \label{fig:peek_pipe}
\end{center}
\end{figure*}

In this section, we describe how the Personalised Educational Engagement linked to Knowledge topics (PEEK) dataset is constructed. \figurename{ \ref{fig:peek_pipe}} outlines the overall process of creating the PEEK dataset.

\subsection{Data Source} \label{sec:data_source}

PEEK dataset is constructed using video metadata and learner activity data extracted from VideoLectures.Net\footnote{\url{www.videolectures.net}} (VLN), a repository of scientific and educational video lectures. VLN repository records research talks and presentations from numerous academic venues accompanied by the lecture slides associated with the video. As the talks are recorded at peer-reviewed conferences and prestigious research venues, the lectures are reviewed and material is controlled for correctness of knowledge.

\begin{figure*}[ht] 
\begin{center}
    \centerline{\includegraphics[width=.7\linewidth]{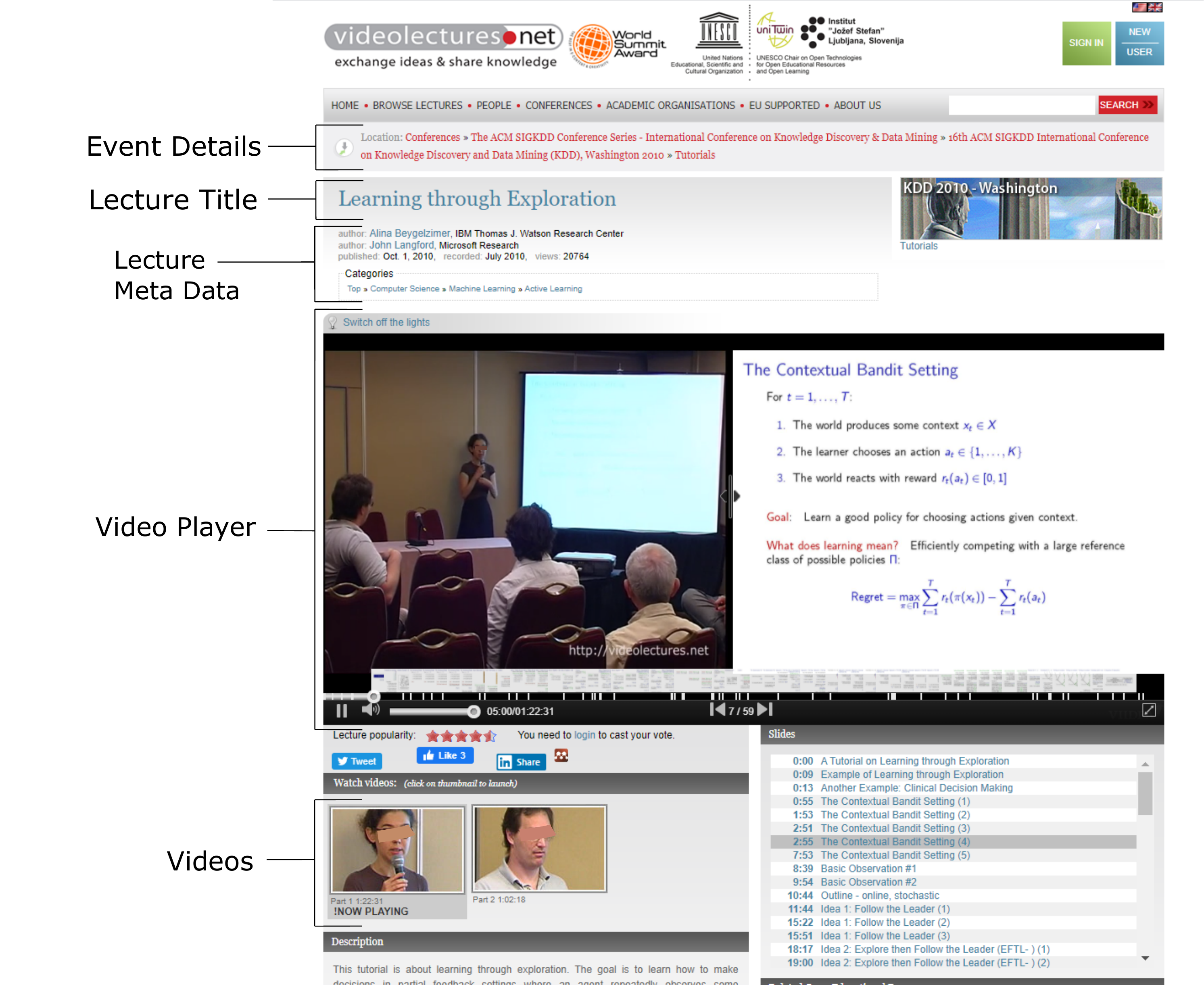}}
    \caption{Screen layout of VideoLecture.Net website from which the data for PEEK dataset was sourced. Every lecture can have one or more videos and is associated with metadata. The lecture transcripts are also available for processing.}
    \label{fig:vln_interface}
\end{center}
\end{figure*}

\figurename{ \ref{fig:vln_interface}} depicts the different components of the VLN user interface where a user engages with an educational video. Every lecture has a title, event details (e.g. conference venue, date, location etc.) and lecture meta data associated with it. Although most lectures consist of one video, some video lectures may contain more than one video as shown in \figurename{ \ref{fig:vln_interface}}. 

\subsection{Fragmenting Videos}

First, the videos in VLN repository are transcribed to its native language using the \emph{TransLectures} project\footnote{\url{www.translectures.eu}}. Then the non-English lecture videos are translated to English as we will use English Wikipedia for entity linking. 

As described in Section \ref{sec:data_source}, a lecture in VLN repository can contain one or more videos. To model user interactions in a much more granular level we further break each video into smaller parts. 
Once the transcription/translation is complete, we partition the transcript of each video into multiple \emph{fragments} where each fragment covers approximately 5 minutes of lecture time.
Having 5 minute fragments allow us to break the contents of a video into a more granular level while making sure that there is sufficient amounts of information in each video fragment.

\subsection{Wikification of Transcripts}
In order to identify the Knowledge Components (KCs) that are contained in different video fragments, we use Wikification\cite{wikifier}. This allows annotating learning materials with humanly interpretable KCs (Wikipedia concepts) at scale with minimum human-expert intervention. This setup will make sure that recommendation strategies build on this dataset will be technologically feasible for web-scale e-learning systems. Previous works have demonstrated the value of using Wikification in this task\cite{truelearn}. We restrict the dataset to English Wikipedia due to its richness in comparison to other languages.  

\begin{figure}[ht]
\begin{center}
    \centerline{\includegraphics[width=\columnwidth]{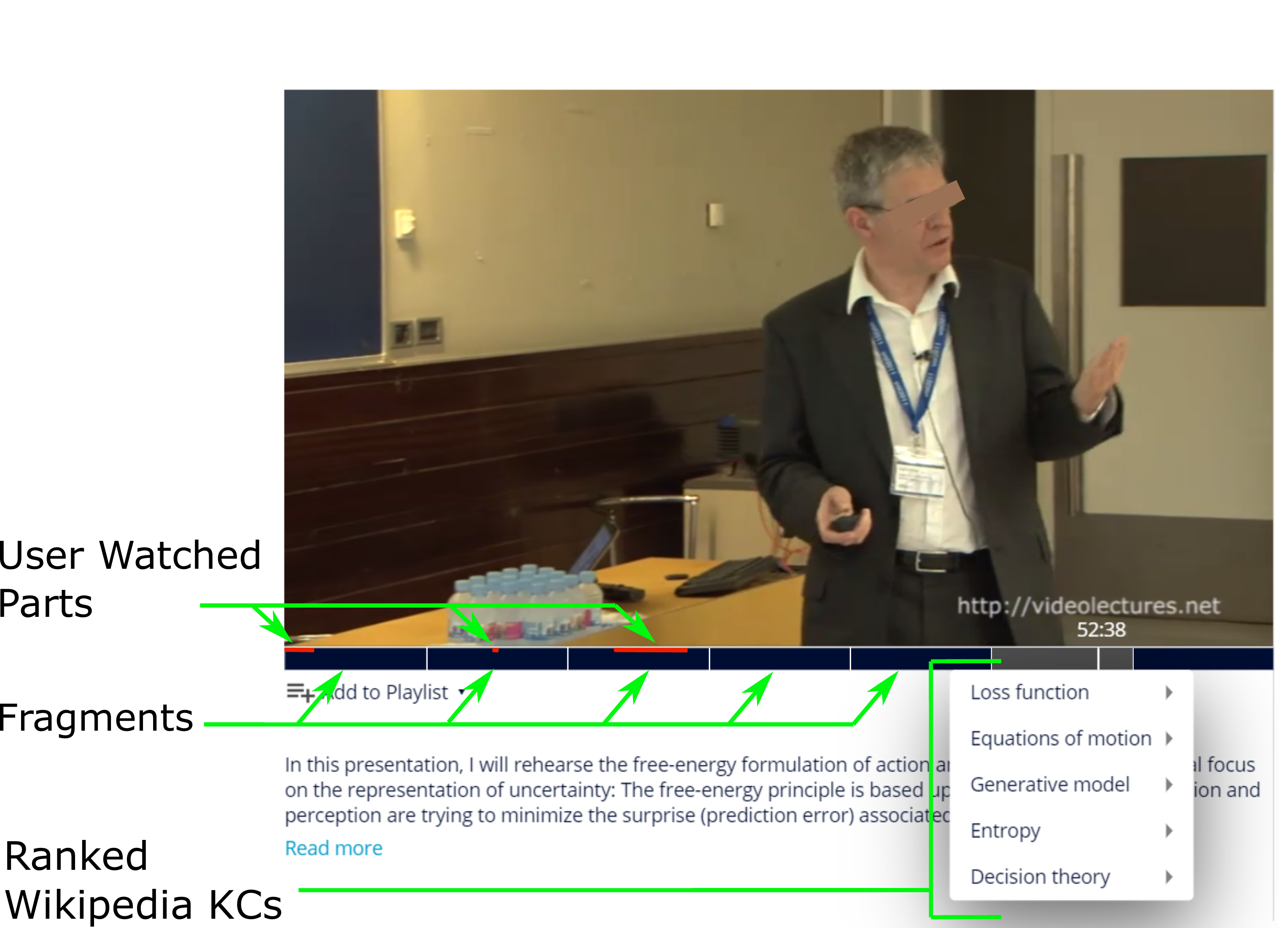}}
    \caption{Visual representation of the different data items available in the PEEK dataset. Each video is broken into multiple, non-overlapping 5 minute fragments that are linked with ranked Wikipedia-based KCs. The watched parts of the video (in red) are used to create discrete engagement labels.}
    \label{fig:fragments}
\end{center}
\end{figure}

\subsection{Knowledge Component Ranking}

As per \cite{wikifier}, Wikification produces two statistical values per annotated KC, namely, \emph{PageRank} and \emph{Cosine Similarity} scores. 

\emph{PageRank score} is calculated by constructing a semantic graph where semantic relatedness ($SR(c,c')$) between each Wikipedia concept pair $c$ and $c'$ in the graph are calculated using equation \ref{eq:wiki_sr}.
\begin{equation}\label{eq:wiki_sr}
        SR(c,c') = \frac{\log(max(|L_c|, |L_{c'}|) - \log(|L_c \cap L_{c'}|)}
        {\log |W| - log (min(|L_c|, |L_{c'}|)}
\end{equation}
where $L_{c}$ represents the set of Wiki concepts with inwards links to Wikipedia concept $c$, $|\cdot|$ represents the cardinality of the set and $W$ represents the set of all Wikipedia topics. This  semantic relatedness graph is used for computing PageRank scores. 
PageRank algorithm\cite{pagerank} leads to heavily connected Wikipedia topics (i.e. more semantically related) within the lecture to get a higher score.

The \emph{Cosine Similarity score} is used as a proxy for topic coverage within the lecture fragment\cite{truelearn}. This score $cos(s_{tr}, c)$ between the \emph{Term Frequency-Inverse Document Frequency (TF-IDF)} representations of the lecture transcript $s_{tr}$ and the Wikipedia page $c$ is calculated based on equation \ref{eq:wiki_cos}:
\begin{equation}\label{eq:wiki_cos}
        cos(s_{tr}, c) = \frac{\texttt{TFIDF}(s_{tr}) \cdot \texttt{TFIDF}(c)}
        {\|\texttt{TFIDF}(s_{tr})\| \times \|\texttt{TFIDF}(c)\|}
\end{equation}
where $\texttt{TFIDF}(s)$ returns the TF-IDF vector of the string $s$ while $||\cdot||$ represents the norm of the TF-IDF vector.

The authors of \cite{wikifier} comment that a linearly weighted sum between the PageRank and Cosine score can be used to rank the importance of Wikipedia concepts as per equation \ref{eq:cos_pr}. 

\begin{equation}\label{eq:cos_pr}
        \texttt{Rank}(c) = \alpha \cdot PageRank(c) + (1 - \alpha) \cdot cos(s_{tr}, c), \text{where } \alpha \in [0,1]
\end{equation}

We experimented with different values for $\alpha$ in equation \ref{eq:cos_pr} and empirically validated suitable linear combinations of weights for PageRank score and Cosine Similarity score. We observed that a weight of 0.8 on PageRank and 0.2 on Cosine similarity leads to the most suitable ranking of KCs\cite{x5gon_d1_3}. We use \texttt{Rank} to identify the five top-ranked KCs for each lecture fragment. The cosine similarity score as per equation \ref{eq:wiki_cos} is included in the dataset as a proxy for coverage of that KC in the lecture fragment. We restrict the number of KCs to 5 as larger numbers (e.g. 10 KCs) have shown to degrade performance of learner models built with similar datasets\cite{truelearn}. \figurename{ \ref{fig:dataset_stats}(ii)} provides a word cloud of the most dominant KCs in PEEK dataset. It is evident that the majority of KCs (Wikipedia topics) associated to the lecture fragments in this dataset are related to artificial intelligence and machine learning.

\subsection{Anonymity}

We restrict the final dataset to lectures that have been viewed by at least 5 unique users to preserve k-anonymity\cite{orcas_dataset} of users. Also, we report the timestamp of user view events in relation to the earliest event found in the dataset obfuscating the actual timestamp. That is we report the smallest timestamp in the dataset $t_0$ as 0s and any timestamp $t_i$ after that as $t_i - t_0$. This allows us to publish the true order and the real differences of duration between events without revealing the actual timestamps. Additionally, the lecture metadata such as title and authors etc. are not published to preserve the anonymity of the authors/lecturers. The motivation behind this decision is to avoid authors of the video lectures having unanticipated effects on their reputation by associating implicit learner engagement values to their content.

\subsection{Labels}
The user interface of VLN website also records the video watching behaviours of its users (Interaction logs in \figurename{ \ref{fig:peek_pipe}}).

The target label for learner engagement is a discrete variable based on \emph{video watch time} which has been used as a proxy for video engagement in both non-educational\cite{Covington2016,beyondviews} and educational\cite{Guo_vid_prod,truelearn} contexts. Normalised learner watch time $\overline{e}^t_{\ell,r}$ of learner $\ell$ with video fragment resource $r_i$ at time point $t$ is calculated as per equation \ref{eq:norm_engage}.

\begin{equation} \label{eq:norm_engage}\centering 
\overline{e}^t_{\ell,r_i} = W(\ell,r_i) / D(r_i),
\end{equation}

where $\overline{e}^t_{\ell,r_i} \in \{0,1\}$, $W(\cdot)$ is a function that returns the \emph{watch time} of learner $\ell$ for resource $r_i$ and $D(\cdot)$ is a function that returns the duration of lecture fragment $r_i$. The ultimate label $e^t_{\ell,r_i}$ is derived by discretising $\overline{e}^t_{\ell,r_i}$ where $e^t_{\ell,r_i} = 1 $ when $\overline{e}^t_{\ell,r_i} \geq .75$ and $e^t_{\ell,r_i} = 0$ otherwise. The discretisation rule is motivated by the hypothesis that a learner should watch approximately 4 minutes of an educational video fragment that is approximately 5 minutes (duration of a video fragment that includes 5000 characters from the video transcript) in order to acquire knowledge from it\cite{truelearn}.

\subsection{Final Dataset}

\begin{figure}[ht]
\begin{center}
    \centerline{\includegraphics[width=.9\columnwidth]{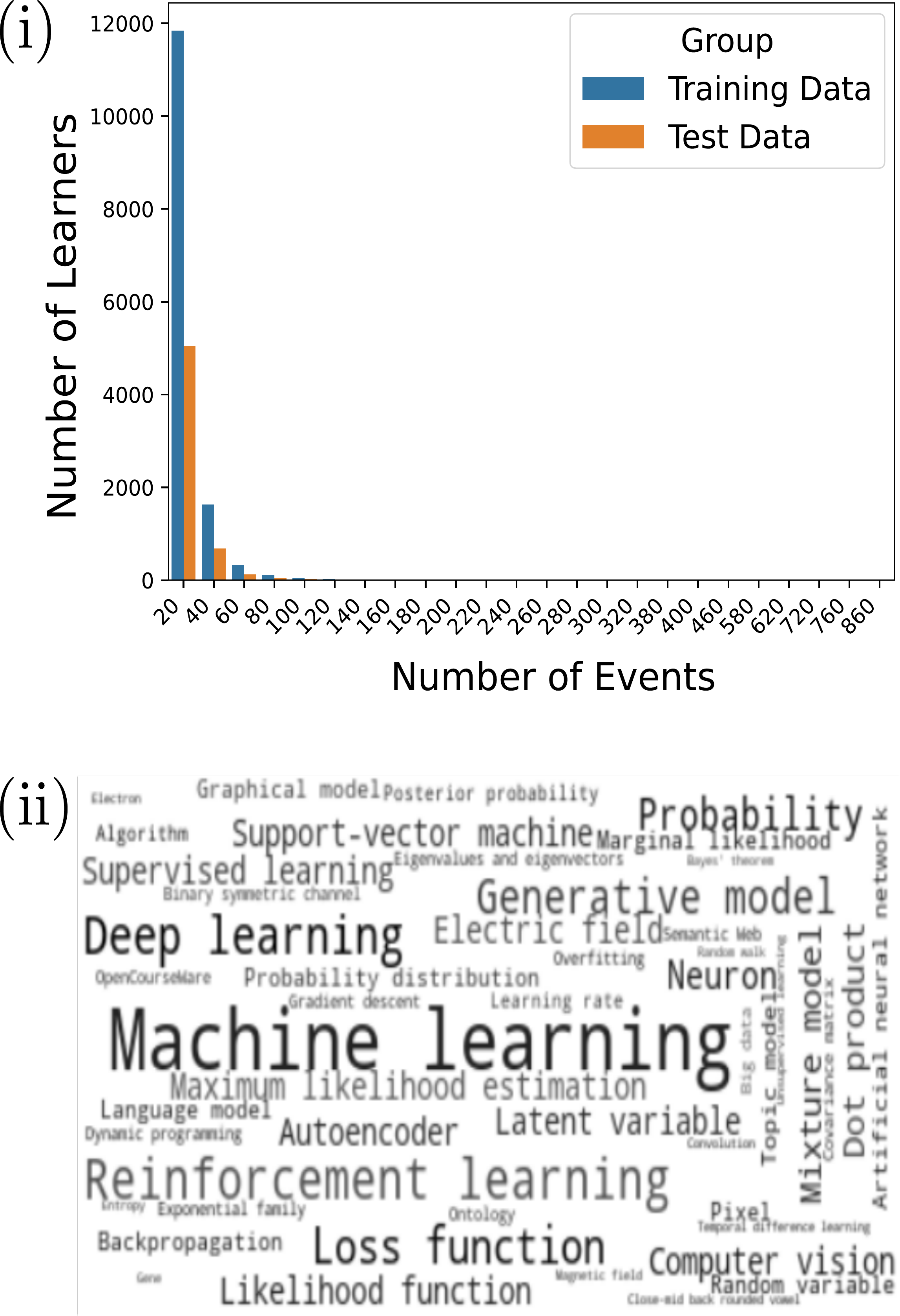}}
    \caption{Characteristics of the PEEK dataset: (i) number of learners in the training/test dataset based on the number of events in the session  for individual learner and (ii) wordclouds depicting the most frequently detected Wikipedia-based knowledge components.}
    \label{fig:dataset_stats}
\end{center}
\end{figure}

The final PEEK dataset consists of 290,535 interaction events from 20,019 distinct users with at least 5 events. These learner engage with 10,233 unique lecture videos that are partitioned into 39,113 fragments (3.82 fragments per video). The learner population in the dataset is divided into \emph{Training} (14,050 learners) and \emph{Test} (5,969 learners) datasets based on a 70:30 split. The label distribution in the dataset is also relatively balanced with only 56\% of the labels being positive. As shown in \figurename{ \ref{fig:dataset_stats} (i)}, the majority of learners in the dataset have a relatively small number of events (under 80) making this dataset an excellent test bed for personalisation models designed to work in data scarce environments. VLN repository mainly publishes videos relating to Computer Science and Machine Learning leading to a learner audience who visit to learn about these subjects. This fact is confirmed by \figurename{ \ref{fig:dataset_stats}} where it shows that the dataset is dominated by events with AI and ML related KCs.

\subsection{Structure of the PEEK Dataset}

The final dataset consists of 3 files. 
\begin{enumerate}
    \item \texttt{train.csv}, used for hyperparameter validation and training parameters.
    \item \texttt{test.csv}, used as the held-out test set.
    \item \texttt{id\_url\_mapping.csv}, which contains the mapping between KC IDs and Wikipedia page URL.
\end{enumerate}
    
\texttt{train.csv} and \texttt{test.csv} files contain the actual learner session data where their interaction with lecture fragments are recorded. The two files contains 70\% and 30\% of the learners respectively. Both files contain 15 columns that are described in Table \ref{tab:features}. As the name suggests \texttt{id\_url\_mapping.csv} file contains a mapping between the KC ID in PEEK dataset and the URL of the Wikipedia page of the concept associated with that KC.

\begin{table*}[] 
\caption{Detailed descriptions of the different columns of the \texttt{train.csv} and \texttt{test.csv} files included in the PEEK Dataset.}
\label{tab:features}
\centering 
\begin{tabular}{ccl}
\toprule
Column Number & Description & Details \\
\midrule
1 & Video Lecture ID & An integer ID associated with an individual video lecture \\
2 & Video ID & An integer ID associated to every video belonging to the same Video Lecture ID (e.g. $1\dots v$ if \\
& &  the lecture with $v$ videos) \\
3 & Part ID & An integer ID associated with each video fragment (e.g $1 \dots f$ for a video with $f$ fragments) \\
4 & Timestamp & Timestamp (to the nearest second) when the play event was initiated. \\
5 & user ID & An integer ID associated with each unique learner in the dataset PEEK dataset (IDs in \\
& & \texttt{train.csv} and \texttt{test.csv} files are mutually exclusive). \\
6,8,10,12,14 & KC IDs & An integer ID associated with each unique Knowledge Component in the dataset. This ID can \\
& & be linked to the human readable Wikipedia concept names found in \texttt{url\_id\_mapping.csv} file \\
7,9,11,13,15 & Topic Coverage & Proxy for coverage of the relevant KC in the fragment of interest. KC   coverage is the cosine \\ 
& & similarity as per equation \ref{eq:wiki_cos}. \\
16 & Label & The binary label $e^t_{\ell,r_i}$, 1 if the learner watched $\geq$ .75 of the video  fragment, 0 otherwise.\\
\bottomrule
\end{tabular}
\end{table*}

\section{Main Prediction Task and Models} \label{sec:exp_and_results}

We benchmark the predictive performance of this dataset with respect to learner engagement prediction by proposing a set of baseline models and measuring their performance with the PEEK dataset. 

\subsection{Data}

Unlike previous studies, which relied on small, expert-curated collections of open educational materials, our dataset is built using in-the-wild watch patterns of multi-lingual open educational materials available online.

{We construct two distinct datasets for our experiments. (i) \texttt{Active 20} is a smaller dataset that consists of 8,207 interaction events from the 20 most active users (410.35 events per learner) in the PEEK dataset. This dataset contains a substantially small number of users who have relatively longer sessions. The smaller dataset will allow us to understand how the proposed learner models will behave with mature learners that have more interaction information in the system. On the contrary, (ii) \texttt{PEEK} is the full dataset that consists of 290,535 interaction events from 20,019 distinct users with at least 5 events. However, the majority of users in this dataset have relatively smaller sessions (average of 14.51 events per learner).} 

\subsection{Prediction Task and Evaluation Metrics}

As the dataset is already divided into train and test splits, all our experiments use a hold-out validation (train-test split) approach  where 
hyperparameters and parameters are learned on 70\% of the learners and the model is evaluated on the remaining 30\% with the selected hyperparameter combination. 

A sequential experimental design is employed, where $\hat{e}^t_{\ell,r_i}$, engagement with fragment $t$ is predicted using fragments $1$ to $t-1$. Since engagement is binary, predictions for each fragment can be assembled into a confusion matrix, from which we compute well-known binary classification metrics such as precision, recall and F1-measure. We focus on these measures as we are more interested in predictive lecture fragments that the learners are likely to engage with.

We average these metrics per learner and weight each learner according to their amount of activity in the system. We use F1-measure for model comparison as we are interested in improving both precision and recall. 

The primary objective of the experiments is to evaluate how the different attributes of the dataset are useful in modelling engagement of learners. Towards this goal, we run two primary experiments, 1) investigate how different recommendation models perform with predicting engagement, and 2) how the number of KCs impact the prediction model. The benchmark algorithms used for the experiments are outlined in Section \ref{sec:benchmarks}. The results of the experiments are outlined in Section \ref{sec:results}.

\subsection{Benchmark Models} \label{sec:benchmarks} 

PEEK is the first of its kind, a dataset that records in-the-wild  engagement of informal learners with video lecture fragments. Due to the novelty of this dataset, we struggle to find already published baselines, except for the TrueLearn family of algorithms\cite{truelearn}. 
For the sake of comparing its predictive performance,  we also propose a set of baselines that are based on content-based and collaborative filtering.

\paragraph{Content-based Similarity} Content-based filtering can measure the similarity between two items. We compute a similarity value, $sim(r^{t-1}_{\ell,r_i},r^{t}_{\ell,r_i})$ between two consecutive lecture fragments $r^{t-1}_{\ell,r_i}$ and $r^{t}_{\ell,r_i}$ in the learner  $\ell$'s session. We use this similarity value to make an engagement prediction $\hat{e}^t_{\ell,r_i}$ based on equation \ref{eq:cbf_baseline}.

\begin{equation} \label{eq:cbf_baseline}\centering 
    \hat{e}^t_{\ell,r_i} = 
    \left\{ 
      \begin{array}{ c l }
        1                 & \text{\ if\ } sim(r^{t-1}_{\ell,r_i},r^{t}_{\ell,r_i}) \geq threshold \\
        0                 & \text{otherwise}
      \end{array}
    \right.
\end{equation}

In this case, we investigate two similarity measures, namely 1) \emph{Cosine}, 2) \emph{Concept-based Jaccard} and \emph{User-based Jaccard}. When computing cosine similarity, we represent each video fragment using the bag of concepts representation where the concepts are the super set of Wikipedia concepts mentioned in the dataset. The values in this sparse vector are the cosine similarities between respective Wikipedia concept and the lecture fragment transcript as per equation \ref{eq:wiki_cos}.

An alternative approach to finding concept-wise similarity is Jaccard similarity. Concept-based Jaccard similarity $\texttt{Jaccard}_{\mathcal{C}}(r^{t-1}_{\ell,r_i}, r^{t}_{\ell,r_i})$,  between lecture fragments $r^{t-1}_{\ell,r_i}$ and $r^{t}_{\ell,r_i}$ is computed based on equation \ref{eq:concept_jaccard}.

\begin{equation} \label{eq:concept_jaccard}\centering 
    \texttt{Jaccard}_{\mathcal{C}}(r^{t-1}_{\ell,r_i}, r^{t}_{\ell,r_i}) = 
    \frac{\mathcal{C}(r^{t-1}_{\ell,r_i}) \cap \mathcal{C}(r^{t}_{\ell,r_i})}
    {\mathcal{C}(r^{t-1}_{\ell,r_i}) \cup \mathcal{C}(r^{t}_{\ell,r_i})}
\end{equation}  
where $\mathcal{C}(\cdot)$ is a function that returns the set of Wikipedia concepts in resource $r_i$

Similarly, one can also measure the similarity between two lecture fragments based on how many learners interact with both the lecture fragments. The user interactions in the training dataset is used exclusively to learn the similarity matrix in order to avoid data leakage. In this approach, we can calculate the user-wise jaccard similarity $\texttt{Jaccard}_\mathcal{U}(r^{t-1}_{\ell,r_i}, r^{t}_{\ell,r_i})$, as per equation \ref{eq:user_jaccard}. 

\begin{equation} \label{eq:user_jaccard}\centering 
    \texttt{Jaccard}_\mathcal{U}(r^{t-1}_{\ell,r_i}, r^{t}_{\ell,r_i}) = 
    \frac{\mathcal{U}(r^{t-1}_{\ell,r_i}) \cap \mathcal{U}(r^{t}_{\ell,r_i})}
    {\mathcal{U}(r^{t-1}_{\ell,r_i}) \cup \mathcal{U}(r^{t}_{\ell,r_i})}
\end{equation}  
where $\mathcal{U}(\cdot)$ is a function that returns the set of learners that interacted with resource $r_i$

\paragraph{Knowledge Tracing (KT)}  

KT builds a learner representation of knowledge of the learner\cite{Yudelson13}. This learning model is then used in predicting engagement of learner $\ell$ with lecture fragment resource $r_i$ at time $t$. As the PEEK dataset has a temporal dimension, we reformulate the KT algorithm into an online learning graphical model inspired by the reformulation in\cite{bishopsnewbook}.  The skill variables in the KT model are Bernoulli variables ($\theta^t_{\ell,c} \sim \texttt{Bernoulli}(\pi^t_{\ell,c})$), assuming that a learner $\ell$ would have either mastered a skill/ concept $c$ or not (represented by probability $\pi^t_{\ell,c}$). Skills are initialised ($\theta^0_{\ell,c}$) using a $\texttt{Bernoulli}(.0)$ prior, assuming that the latent skill is not mastered in the beginning. A noise factor similar to what is found in the conventional KT model\cite{corbett1994knowledge} is added to this model and is tuned using a grid search.
    
\paragraph{TrueLearn Novel Model}
Similar to KT model, TrueLearn model is also an online, graphical model that develops a learner model of the learner. This model is inspired by the TrueSkill model\cite{trueskill} which is an evolution of IRT\cite{Rasch1960}. Contrary to the KT model, TrueLearn model models skills as Gaussian variables ($\theta^t_{\ell,c} \sim \mathcal{N}(\mu, \sigma^2)$. In addition to modelling knowledge, TrueLearn Novel model also models novelty of content which is a key aspect of educational recommendation\cite{truelearn}. Enforcing the same assumptions of KT model, the TrueLearn Novel skill parameters are initialised using a $\mathcal{N}(.0, \sigma_0^2)$ prior where $\sigma_0^2$, initial variance is a hyperparameter tuned using a grid search. 


\section{Prediction Task Performance and Discussion} \label{sec:results}
The predictive performance of multiple benchmark model (outlined in Section \ref{sec:benchmarks}) with the PEEK dataset is evaluated. The hyperparameters of each model (including number of KCs) are tuned using a grid search. The model performance with the best hyperparameter combinations is presented in Table \ref{tab:results_overall}.

\begin{table*}[] 
\caption{Predictive performance of suitable }
\label{tab:results_overall}
\centering 
\begin{tabular}{lcccccc}
\toprule
\multicolumn{1}{r}{Dataset} & \multicolumn{3}{c}{Most Active 20 Users} & \multicolumn{3}{c}{PEEK Dataset} \\
Algorithm & \multicolumn{1}{c}{Precision} & \multicolumn{1}{c}{Recall} & \multicolumn{1}{c}{F1-Score} & \multicolumn{1}{c}{Precision} & \multicolumn{1}{c}{Recall} & \multicolumn{1}{c}{F1-Score} \\
\midrule
\texttt{Cosine} Similarity & \textit{67.799} & \textit{87.194} & \textit{73.635} & \textit{57.859} & 58.451 & 54.056 \\
$\texttt{Jaccard}_{\mathcal{C}}$ Similarity & 67.157 & 80.901 & 70.818 & 57.810 & 60.362 & 55.026 \\
$\texttt{Jaccard}_{\mathcal{U}}$ Similarity & 46.499 & 11.606 & 17.761 & 57.854 & \textit{72.755} & \textit{61.219} \\
Knowledge Tracing & 65.050 & 65.828 & 62.077 & 53.254 & 28.564 & 34.513 \\
TrueLearn Novel & \textbf{78.790} & \textbf{89.390} & \textbf{83.090} & \textbf{58.290} & \textbf{79.240} & \textbf{64.710} \\
\bottomrule
\end{tabular}
\end{table*}

Good predictive power of the similarity based approaches validate the expressiveness of the content representations based on Wikipedia concepts and cosine similarity. The poor performance of \texttt{$Jaccard_{\mathcal{U}}$} Similarity model with the top 20 user dataset is expected as this model relies on learning similarities based on training data. Although 20 user dataset contains active users with more events per session, there is less data in total due to the small number of users which leads to learning a mostly incomplete similarity matrix between lecture fragments. It is seen that this condition changes with the full PEEK dataset where more data is available. The results in table \ref{tab:results_overall} shows that TrueLearn Novel model performs best among the benchmarks investigated. This result further reinforces the superiority of TrueLearn Novel model investigated with a similar dataset\cite{truelearn}.
 
\subsection{Number of Knowledge Components}

\begin{figure*}[ht]
\begin{center}
    \centerline{\includegraphics[width=.95\linewidth]{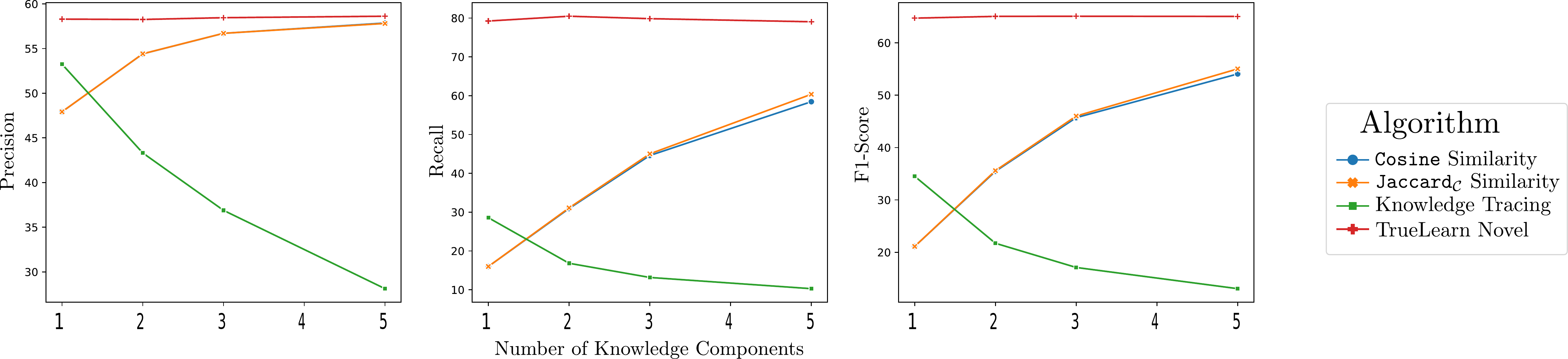}}
    \caption{Predictive performance of PEEK dataset in terms of Precision (left), Recall (middle) and F1-Score (right) for the benchmark models when varying numbers of Knowledge Components (KCs) are used as the content representation.}
    \label{fig:topics}
\end{center}
\end{figure*}

The impact of the number of KCs for each model is also investigated and the results are reported in \figurename{ \ref{fig:topics}}.

The plots indicate that the similarity based approaches tend to perform better with more KCs in the representation. This is natural as Similarity based approaches need more information from the exact pair of contents that are being compared in order to make a fine grained prediction. On the contrary, Knowledge Tracing and TrueLearn novel Models that develop a learner model perform better with less number of KCs. This observation is also expected as the learner model has the capacity to store information about various different KCs that the learner encountered over the past. This allows the learner model to make a good prediction with little information from the current resource. In the TrueLearn model, the effect of adding extra KCs to the representation makes a very small difference in performance.

\subsection{Discussion}

Knowledge (skill mastery to be more specific) is usually measured from explicit test performance\cite{corbett1994knowledge,deep_kt,gkt}. Contrary to narrowly scoped learning scenarios common to ITS, seeking explicit feedback is comparatively less realistic in lifelong learning settings where informal learners attempt to acquire knowledge without necessarily having the need to go through rigorous testing. Forcing learners into aggressive testing can hinder their learning experience. Additionally, technical challenges too exist in areas such as scalable test/ question generation and automatic test marking. 
However,\cite{truelearn} has demonstrated that modelling knowledge from educational watch patterns is a promising direction towards predicting future engagement. The opportunity that PEEK dataset brings is the capability to advance this research avenue by making the first large dataset publicly available. This dataset will allow researchers to improve knowledge representation and also include other crucial factors such as learner interest\cite{trueeducation} in the spirit of building integrative educational recommenders using implicit feedback. As \figurename{ \ref{fig:dataset_stats}(i)} shows, PEEK dataset also contains a large number of learners with limited activity making this dataset an excellent candidate to evaluate low resource, data efficient personalisation algorithms.   

PEEK dataset also acts as the bridge between multiple rich data sources. The Wikipedia concepts used in representing the KCs also enables connecting this dataset to the variety of auxiliary information available around Wikipedia, the world's largest encyclopedia. Some examples of possibilities are the ability to leveraging additional data structures such as Wikipedia page contents, the hyperlink graph and the category tree. Other enriched derivatives of Wikipedia such as semantic relatedness\cite{ponza_semantic_relate} and knowledge bases\cite{auer2007dbpedia,wikidata} are also directly fused with this dataset to build more powerful content representations. In addition, the Wikipedia taxonomy also provides humanly intuitive representations (grounded on Wikipedia concepts) that enhances interpretability. \figurename{ \ref{fig:dataset_stats} (ii)} is a prime example of how humanly intuitive KCs improves the expressiveness of content representations. A humanly intuitive representation also paves way to much needed interpretable learner models that are essential to triggering meta-cognition within learners\cite{bull2008metacognition}.

\subsection{Relevance to Online Learning}
PEEK dataset consists of a collection of learners making engagement choices across educational materials over time. This dataset clearly captures the temporal dynamics of the learners whose knowledge state and preferences change over time. The superiority of TrueLearn Novel in Table \ref{tab:results_overall} also gives a strong indication on how  a state-aware learner model that changes its learner representation over time, best captures the engagement dynamics of the dataset in comparison to na\"ive similarity-based approaches. This result reaffirms the relevance and usefulness of this dataset to invent online learning models that can identify personal and temporal dynamics of users.

\subsection{Supported Tasks} \label{sec:taks}

This section introduces the reader to the tasks that the PEEK dataset could be used for. The main application area of the PEEK dataset is engagement prediction with video lecture fragments in an web-based learning setting as demonstrated in Section \ref{sec:exp_and_results}. However, this task can be further extended by fusing the engagement labels from all the fragments of a video to carry out video recommendation which is more common\cite{Covington2016} (contrary to video fragment recommendation). 

Moving away from engagement prediction, PEEK dataset can also be used to understand the structure of knowledge and learning pathways through it. Deducing the structure of knowledge using the co-occurrence patterns of KCs within the video lectures provide opportunities to understand inter-topic relationships and how knowledge is structured. Work in this direction can be used in identifying related materials and accounting for novelty in educational recommendation\cite{truelearn}. The dataset can also be used to identify different clusters/groups of learners and learning resources which will allow understanding the education landscape better. Studying the evolution of KCs within learner journeys over time can also unlock opportunities to understand prerequisites\cite{yu2020mooccube}. 

\section{Conclusions}
This work releases \emph{PEEK}, the first and largest publicly-available, humanly-intuitive dataset of learner interaction with educational videos opening up various avenues to the research community to push the frontiers of this line of research. It publishes learner interactions of over 20,000 learners with fragments of lectures over time. The predictive performance of several benchmark models is evaluated with this dataset. The power of this dataset is already portrayed by the promising experimental results presented.

As future directions, we see the importance of modelling semantic relatedness between KCs\cite{ponza_semantic_relate} that are likely to boost performance of learner models such as Knowledge Tracing and TrueLearn. Exploiting other side information from other Wikipedia such as the category tree is also an interesting direction to explore. The potential of detecting other user signals such as learner interest from PEEK dataset can also lead to insightful findings. PEEK dataset also allows experimenting with temporal dynamics such as interest decay\cite{10.1145/2993318.2993332}.  

\begin{acks}
This work is partially supported by the European Commission funded project "Humane AI: Toward AI Systems That Augment and Empower Humans by Understanding Us, our Society and the World Around Us" (grant 820437) and the EPSRC Fellowship titled "Task Based Information Retrieval" (grant EP/P024289/1).
\end{acks}

\bibliographystyle{ACM-Reference-Format}
\bibliography{sample-base}

\end{document}